\documentclass[%
 aip,
 jmp,%
 amsmath,amssymb,
 reprint,%
]{revtex4-1}

\usepackage{graphicx}
\usepackage{grffile}
\usepackage{dcolumn}
\usepackage{bm}
\usepackage{multirow}
\usepackage{color} 
\usepackage{etex}
\reserveinserts{58}
\usepackage{morefloats}

\begin{document}

\preprint{XXXXX (preprint)}

\title[Versinus visualization of graphs]{Versinus: a visualization method for graphs in evolution}

\author{Renato Fabbri}%
 \homepage{http://ifsc.usp.br/~fabbri/}
 \email{fabbri@usp.br}
  \affiliation{ 
S\~ao Carlos Institute of Physics, University of S\~ao Paulo (IFSC/USP)
}%
%
%
%
%
%
%
%
%
%
%

\date{\today}

\begin{abstract}
This article presents a novel visualization approach for dynamic graphs, the versinus method,
specially useful for real world networks exhibiting free-scale properties. With a simple and fixed layout, and a small set of visual markups, the method has been useful for understanding network dynamics. Local community often suggests that it be reported, which motivated this article. Online resources deliver videos and computer scripts for rendering new animations. This article has a concise description of the method.
\end{abstract}

\maketitle

%

\section{\label{sec:into}Introduction}
Versinus is a visualization method for dynamic graphs based on experimental observations.
This method receives this dedicated article by recurrence of the suggestion, by fellow researchers,
to write it.

In visualizing a network, the method consists of creating an animation,
of a fixed-size message sliding window (e.g. 400 messages), and 
partitioning the network in two fixed-layout segments:
a sinusoid for the most connected vertexes
and a straight line for the less connected.
A vertex holds the same position throughout the animation. Also,
visual indication of properties - such as color, height and width,
and rank of vertex with degree criteria - plays a central role.
This work differs from the few works on the visualization of dynamic
graphs because it is a simple report on a method that has been useful,
was developed by practical needs, is the result of experimentations,
and has a number of criteria that guided its development~\cite{Viz1,Viz2,Viz3}.

Next section further exposes the context in which versinus is being useful,
and the characteristics of the systems in which it was applied.
Section~\ref{vmethod} describes the method itself. Section~\ref{res} is dedicated
to most important results and discussions. Conclusions and further work
are in Section~\ref{conc}. Appendix exposes online resources.

\section{Context and materials}\label{cmaterials}
In 2013, while doing complex networks research, within a physics department,
to visualize networks undergoing its evolution along time was a necessity.
This posed some challenges (e.g. proper parametrization) to which two
visualization methods were developed. The first is a web gadget, accessible via
usual web browsers, that could render image galleries with
measures and the node-edge structure~\cite{galGMANE, appGMANE}.
The second method 
is based on rendering animations with specific characteristics.
This second method has been named \emph{versinus} for the purpose of this article.

\subsection{GMANE email lists interaction networks}\label{intNet}
The networks undergoing analysis were email interaction networks.
Messages were obtained from the GMANE public database~\cite{GMANE,GMANE2,GMANEwikipedia}. For this goal, just three metadata fields of each email message is needed: the sender, the ID of the message and the ID of the message it is a response to, if any. Each sender is a vertex in the network. If a message is a response, it yields an edge from the author of the original message to current message author. If the edge is already there, the weight is increased by one (an edge is created with weight one). By inspecting all messages in a set, the interaction network given by those messages can be attained. This interaction network is understood as an ``information network'', with edges along the information flow, as the response is an evidence that the responder absorbed the information given by the original sender. If the edges are inverted, it is usual to call it a ``status network'', with the responder "giving status" to the original sender, as he considered the message worthy enough to respond it.

\subsection{Sliding message window dynamic interaction networks}\label{sec:sliding}
Consider $\Delta$ a fixed number of consecutive messages (e.g. $\Delta=400$). Consider sets $s_{i}^{i+\Delta}$ of $\Delta$ consecutive email messages, as they are registered along time. A Sequence $S^{\Delta,M}$ of such sets, with the first message positioned in each the $M$ considered messages (e.g. $M=20000$), can be written as: 

\begin{equation}
S^{\Delta,M}=\{s_i^{i+\Delta}\}_{i=0}^{M-\Delta}
\end{equation}

Each set $s_i$ yields an interaction network, as described in Section~\ref{intNet}. Each of these sets exhibits stable properties, while each participant exhibits a wide variation of characteristics along $S$. To understand the mechanisms of this compatibility (unstable vertexes, stable network) led to experimenting a series of layouts and visualization techniques, from which versinus emerged.

\section{The versinus visualization method}\label{vmethod}
\begin{figure}[h!]
    \begin{center}
        \includegraphics[scale=.25]{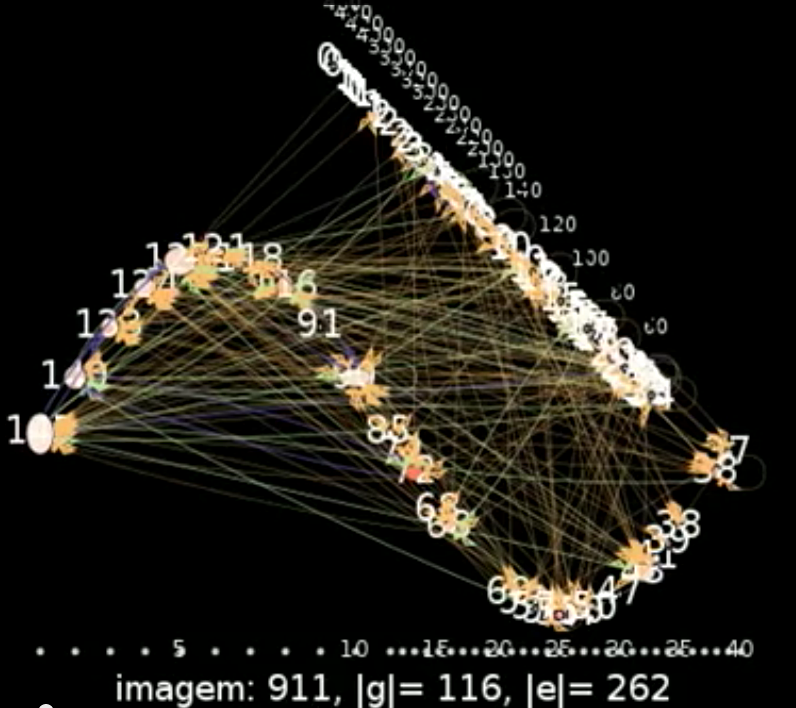}
        \caption{The versinus visualization method in use. 5\% of The most connected vertexes (hubs) are on the left period of the sinusoid. 15\% of the most connected remaining vertexes are on the right period. 80\% of the least connected vertexes are on the straight line, above the sinusoidal shape. White dots with numbers keep track of node position in the overall degree ordering. Measures blink periodically on the vertex it is related to.}
        \label{fig:versinus}
    \end{center}
\end{figure}

Versinus is a simple visualization method. Network vertexes are roughly split to usual $80\%$ periphery, $15\%$ intermediary and $5\%$ of hubs. Hubs are laid on the first half of a sinusoid. Intermediary on the second. Peripheral sector is laid on a straight line. This configuration can be improved in various forms, to which Section~\ref{sec:ref} is dedicated. Figure~\ref{fig:versinus} has an image of such a layout. The position of vertex, fixed, is defined by the overall structure, i.e. with respect to all $M$ messages (see Section~\ref{sec:sliding}). Numbers with individual measures for each vertex blink periodically.

\section{Results and discussion}\label{res}
Results from versinus is divided in two groups:
observations on features that made it useful for the task,
and the network structure properties it made possible to grasp.

\subsection{Understanding of useful visualization features for dynamic networks}

On the numerous insights related to versinus, a few of them seem more fundamental, or plain useful. This is an attempt to present them in an importance-first order:
\begin{enumerate}
    \item Vertexes need to remain static. Even if they move smoothly, one notices solely transient artifacts if nodes are moving around.
    \item Very connected sectors (hubs and intermediary) need to be laid in a curve, otherwise the edges enclose each other and reasoning about the network becomes harsh.
    \item Height and width of a vertex are very informative, specially if measures mapped to them have a strong relation, such as out-degree (mapped to height in versinus) and in-degree (mapped to width).
    \item Coloring nodes was also informative, although less than height and weight, as these are related to measure differences more directly.
    \item An ordering of nodes, related to their fixed position, is very useful. Among all tests, ordering of vertexes by degree was considered the most informative order, which led to the hub, intermediary and peripheral sectioning. As node position in layout is fixed, its ordering is done with respect to the overall structure, i.e. considering all $M$ messages.
    \item  Numbering these positions with respect to the order of the vertexes in the larger structure (all $M$ messages) is useful in understanding how much a vertex holds his position in his scale and on a larger window size.
\end{enumerate}

Many other insights were given by versinus, such as possible visualization tools, other kind of convenient layouts and glyph elaborations. These received dedicated attention in Section~\ref{sec:ref}.
Even so, those insights, numbered above, were incorporated to versinus as the result of tests which presented clear benefits within the best settings achieved.

\subsection{Understanding of network properties through versinus}
A number of hypothesis were drawn about networks, for which versinus
was designed. Another number of hypothesis were driven from versinus use itself.

\subsubsection{Hypothesis evidenced by versinus}
As suggested by Palla, Barab\'asi and Vicsek~\cite{barabasiEvo}, stability of participant activity in social networks is more incident in smaller networks. In accordance with this result, all hubs have intermittent activity in the settings analyzed, except for the email list with the smallest number of participants (the Metareciclagem email list). The intermitence of hubs itself was one of the top hypothesis which motivated versinus development. The stability of the structure~\cite{evoSN} required further verification of the instability of each participant activity.

\subsubsection{Hypothesis drawn from versinus use}
There seem to be modes of operation of the network. As an example, the intermediary is observed to communicate mostly to hubs or to communicate substantially with the peripheral vertex. Another hypothesis is that in and out degree discrepancies in each vertex are important for network characterization. This was confirmed by a recent work by the present author~\cite{evoSN}. Other hypothesis, such as discrepancies in authority and connectivity of vertex, are numerous but need further research to be valuable.

\section{Conclusions and further work}\label{conc}
Versinus has been calling the attention of fellow
researchers for its simplicity and utility. This concise
exposition of the method is aimed at giving further
consequence for this enthusiasm. Results are satisfactory
in the domain of interest and applications of the method
to visualize other domain dynamic networks should come in near future.

\subsection{Refinements on versinus}\label{sec:ref}
By far, the most proficuous facet of versinus is its ability
to bring insights about how to enhance its layout and use.
Besides the technological proposal of making a tool for using versinus
in real-time, there are some ideas about the layout and visual guides itself. 

To further enable visualization of hubs and intermediary vertex,
the sinusoid can have many periods 
with a decaying frequency.
The upper straight line can also have an oscillating trace.
The two halves of the sinusoidal period can be moved independently.
The waveform need not to be a sinusoid.
One can think of many ways to make more informative glyphs. Also, visual and auditory signals for specific occurrences can be interesting (e.g. when a new vertex appears, when one vanishes, when an ordering of vertexes changes).
Measures of each vertex can be exposed with a vertical displacement, to enable multiple measures, to avoid the necessity to blink the numbers and to keep network visualization free from occlusion.

\subsection{Other strategies envisioned}\label{sec:other}
Working with versinus has suggested other kinds of layout for vertexes, specially other geometric figures and iterative force-based methods for positioning vertex in a fixed layout. Matrix visualization of graphs and recent approaches~\cite{Viz1} has been foreseen as support to versinus.

\appendix

\section{Online videos}
There are video playlists online for use by the research team. Authors
choose~\cite{animacoes} as an example. Other playlists can be found by inspecting the user.

\section{Using online scripts for rendering versinus animations}
For making the animations and the analysis of the networks, Python scripts are online in a public repository~\cite{scriptsFim}.

\vspace{5mm}
\nocite{*}
\bibliography{paper}

\end{document}